\documentclass[showpacs,preprint,amsmath,amssymb,aps,12pt]{revtex4}

\usepackage{graphicx,rotating,lscape}
\usepackage{bm}

\begin{document}

\def\erf{\mathrm{erf}}

\title{Latent heat of  nuclear matter}
\date{\today}

\author{Arianna Carbone and Artur Polls}
\affiliation{Departament d'Estructura i Constituents de la Mat\`eria and Institut de 
Ci\`{e}ncies del Cosmos, Universitat de Barcelona, Avda. Diagonal 647, E-08028 Barcelona, Spain}

\author{Arnau Rios}
\affiliation{Department of Physics, University of Surrey, Guildford, Surrey, GU2 7XH, United Kingdom}

\author{Isaac Vida\~na}
\affiliation{Centro de F\'{i}sica Computacional, Department of Physics, University of Coimbra, 
PT-3004-516 Coimbra, Portugal}

\begin{abstract}

We study the latent heat of the liquid-gas phase transition in symmetric nuclear matter using  
self-consistent mean-field calculations with a few Skyrme forces. The temperature
dependence of the latent heat is rather independent of the mean-field parametrization
and it can be characterized by a few parameters. 
At low temperatures, the latent heat tends to the saturation energy. Near the critical 
point, the latent heat goes to zero with a well-determined mean-field critical exponent.  
A maximum value of the latent heat in the range $l \sim 25-30$ MeV is found at intermediate
temperatures, which might have experimental relevance.
All these features can be explained from very basic principles.

\end{abstract}


\maketitle

\section{Introduction}

The study of the liquid-gas phase transition in homogeneous symmetric nuclear matter 
provides interesting links between statistical mechanics, quantum physics and nuclear 
dynamics \cite{siemens83,tsang06,borderie08}. 
The liquid-gas phase transition picture can, in principle, be applied to nuclear systems 
\cite{pochodzalla97}. The nucleon-nucleon (NN) interaction has a qualitative similar 
structure to the inter-atomic one:
it is repulsive at short relative distances and attractive at intermediate and long
distances \cite{RiSch}. At zero temperature, this structure drives the particles to stay at a 
given distance from each other, thus leading to a structured liquid phase. 
As temperature increases, 
thermal fluctuations overcome the interaction effects and the system tends to gasify, thus 
causing a phase transition \cite{huang}.

The only way to study thermodynamical properties of nuclear systems on Earth is via
intermediate and heavy ion collisions  \cite{siemens83,pochodzalla97}, which 
unfortunately are particularly fast ($\sim 10^{-21}$s)  processes and very difficult
to model. Unlike in other branches
of physics, one cannot define or prepare samples for a thermodynamical experiment.
Furthermore, in a collision the state of the system changes continuously in time,
and it is difficult to establish whether the system has actually reached a state of 
equilibrium or not \cite{hauger96,viola99}.  
Finally, the fact that we are dealing with finite nuclear matter
causes additional difficulties on the thermodynamical interpretation of the 
results \cite{bonche84}.

Experimentally, the challenge lies in being able to control the equilibrium, \emph
{i.e.} in determining the state variables of the system, like the temperature, the pressure or the 
density of the system \cite{natowitz95,serfling98}. Moreover, one also needs to find 
suitable observables that help to identify the 
liquid-gas phase transition. Candidates
include, among others, the critical behavior of fragment partitions 
\cite{campi88,gilkes94,elliott02}, 
energy fluctuations on an event by event basis \cite{dagostino00}, 
nuclear calorimetry \cite{pochodzalla95,natowitz02b}
and bimodality of the largest fragment distribution \cite{pichon08,bonnet09}.
The evidence gathered with these different experimental techniques points towards 
the existence of a liquid-gas phase transition for nuclear systems at densities 
below the empirical saturation density, $\rho_0=0.16$fm$^{-3}$, and temperatures 
around $~5-10$ MeV.

From the theoretical side, symmetric nuclear matter at finite temperature provides a first 
qualitative picture of the thermodynamics of nuclear systems and, particularly, of the
liquid-gas phase transition \cite{tsang06,borderie08}. Because of its relative simplicity, 
nuclear matter has been the 
subject of numerous investigations. From a microscopic perspective, different many-body 
techniques 
(including Dirac-Brueckner \cite{terhaar86} and Brueckner--Hartree--Fock 
\cite{baldo99}, Self-Consistent Green's functions \cite{rios08,soma09} or variational 
calculations \cite{mukherjee07}) have been used to compute the thermodynamical 
properties of nuclear matter starting from basic nucleonic degrees of freedom and 
phase-shift equivalent realistic NN interactions.  
So far, very different predictions for critical properties have been obtained using different
approaches or even different interactions within the same approach \cite{rios08}.
A more efficient way to get insight in the thermal behaviour of nuclear matter is by means of 
effective interactions in the framework of a Hartree-Fock approximation, either relativistic
\cite{muller95} or non-relativistic \cite{vautherin96,sil04}. 
Mean-field calculations are much faster to implement numerically and give access
to a wider range of phenomena (critical exponents, for instance). 
Moreover, the mean-field parametrizations
are directly connected to nuclear structure \cite{bender03}.
Recently, one of us has studied systematically the dependence on the effective 
interaction of the liquid-gas phase transition of nuclear matter with Skyrme and Gogny
mean-fields \cite{rios10}. 

In the context of homogeneous nuclear matter, one generally assumes that 
thermodynamical equilibrium takes place between two infinite pieces 
of nuclear matter, one of them belonging to the liquid phase and the other to a gaseous
state. This is quite a crude approximation that can only provide an average, qualitative 
understanding of the thermal  properties of the liquid-gas phase transition.
As a matter of fact, one should consider that the low-density phase of dense nuclear 
matter is not necessarily homogeneous \cite{horowitz09}.
Its description should allow also for the formation of light clusters of nucleons, including
deuterons, tritons, helions ($^3$He) or $\alpha$-particles \cite{goodman84,beyer00,typel10,avancini10}. 
Moreover, finite size and Coulomb effects are very important to determine the
thermal properties and the critical behavior of finite fermionic systems \cite{gulminelli03}.

Very little attention has been paid so far to the study of the latent heat of nuclear matter, 
which is the main goal of this work. This is a very interesting quantity, since it can 
provide a further characterization of the liquid-gas phase transition, and it can be 
potentially extracted
both from experiments and theory. Experimentally, the latent heat might be, in 
principle, read out from
the length of the plateau in the caloric curve (\emph{i.e.} in the temperature vs. excitation 
energy curve). This would suggest values of $l \sim 4-8$ MeV
\cite{pochodzalla95,lee97,natowitz02b}. Note, however, that this plateau can also be 
explained in other terms, rather than in a phase 
transition picture \cite{campi96,sobotka04,shetty09}. Alternatively, the latent heat can also 
be obtained from  the heat capacity versus excitation energy, as shown in Ref.~\cite
{dagostino00}, in which case $l \sim 2-4$ MeV. Recently, bimodality in the charge 
distribution of the heaviest fragments has been used to extract a 
latent heat value of $l \sim 8$ MeV \cite{bonnet09}.
In any case, one has to be aware of
the fact that, for the latent heat to be defined, 
the reaction employed to extract it should not be an isentropic process \cite{randrup09}.

In the case of infinite nuclear matter, the latent heat per particle accounts for the amount of heat 
needed to take a nucleon from the liquid to the gas phase. In this work, we want to 
study this quantity and provide estimates for its typical values in nuclear matter. 
We will, in particular, describe its basic temperature dependence with different 
Skyrme mean-field parametrizations.
We also want to link the behavior of this quantity to the properties of the 
underlying NN interaction. Ultimately, our aim is to understand the behavior of the 
latent heat from basic 
quantum statistical mechanics arguments.

In the next section, we discuss very briefly the formalism needed for the calculation 
of the latent heat within the Skyrme-Hartree-Fock approximation. 
After that, in Section III, we present and discuss the self-consistent results, 
which is followed by 
a section devoted to the analysis of the low temperature and critical behavior of the latent heat.
A summary and final conclusions are presented in the last section. 

\section{Formalism}

To describe efficiently finite nuclei and 
nuclear matter at the same time, one generally relies on phenomenological interactions, 
adequate for the Hartee-Fock (HF) approximation. The effective forces used in
the following are of the Skyrme type. They were introduced
in the 1950's \cite{Skyrme1,Skyrme2}, and have been intensively used in the
literature \cite{stone03}. Alternatively, one can obtain equivalent results by
formulating the problem in terms of density functionals \cite{gupta80,bender03}. 

In this work we deal with a  Skyrme effective interaction:
      \begin{eqnarray}
      \hat{v}_{ij}&=&t_0(1+x_0P_\sigma)\delta(\overrightarrow{r})
      +\frac{1}{2}t_1(1+x_1P_\sigma)[\delta(\overrightarrow{r}) \overrightarrow{k}^2+
      \overleftarrow{k}^2\delta(\overrightarrow{r})]
      \nonumber \\
      &+&t_2(1+x_2P_\sigma)\overleftarrow{k}\delta(\overrightarrow{r})\overrightarrow{k}
       +\frac{1}{6}t_3(1+x_3P_\sigma)
      [\rho(\vec{R})]^\alpha\delta(\overrightarrow{r})\, ,
      \label{eq:skyrme}
      \end{eqnarray}
where 
$\overrightarrow{R}=(\overrightarrow{r_i}+\overrightarrow{r_j})/2$, 
$\overrightarrow{r}=\overrightarrow{r_i}-\overrightarrow{r_j}$, 
$\overrightarrow{k}=(\overrightarrow{\nabla_i}-\overrightarrow{\nabla_j})/2i$ is acting on the right;
$\overleftarrow k =-(\overleftarrow{\nabla_i}-\overleftarrow{\nabla_j})/2i$,
acting on the left;
$P_\sigma=(1+ \overrightarrow{\sigma_1} \cdot \overrightarrow{\sigma_2})/2$ is the spin-exchange operator 
($\vec{\sigma}$ are Pauli matrices;),
and $\rho=N/\Omega$, the total baryonic density. The parameters
$t_0,t_1,t_2,t_3,x_0,x_1,x_2,x_3,\alpha$ are numerical constants fitted to reproduce,
in general, the saturation properties of nuclear matter and structure properties 
of closed shell nuclei \cite{stone03}. 
Three-body interactions are effectively accounted for by the last density-dependent term. 
Mean-field calculations with effective Skyrme forces reproduce in a 
satisfactory way the structure of a wide range of nuclei \cite{bender03}.

Due to the translational invariance of uniform nuclear matter, single-particle (s.p.)
states are appropriately described by plane waves, and the relevant quantum numbers are
the s.p. momentum, $\vec k$, as well as the spin and isospin projections. 
At finite temperature within the Hartree-Fock approximation, momentum states are occupied
according to the Fermi-Dirac distribution,
      \begin{equation}
      n({\vec k},T)=\frac{1}{1+e^{(\varepsilon({\vec k})-\mu)/T}}\,,
      \label{eq:fddistrib}
      \end{equation}
where $T$ is the temperature of the system; $\mu$, its chemical potential, 
and $\varepsilon({\vec k})$,  the s.p. energy. Including the rearrangement term arising from 
the density dependence of the effective interaction, the latter is written as:
\begin{equation}
\varepsilon({\vec k})=  \frac{\hbar^2 k^2}{2 m^*} + 
\frac{1}{16} {\cal T} \left( 3 t_1 + 5 t_2 + 4 t_2 x_2 \right) 
+ \frac{3}{4} \rho \left[ t_0 + \frac{1}{12} ( \alpha + 2 ) t_3 \rho^{\alpha} \right] \,  ,
\end{equation}
where the effective mass,
\begin{equation}
\frac{m^*}{m}= \frac{1}{ 1 + \frac{2m}{\hbar^2} \frac{1}{16} \left( 3 t_1 + 5 t_2 + 4 t_2 x_2 \right) \rho  } \, ,
\label{eq:effmass}
\end{equation}
and the kinetic energy density,
\begin{align}
	{\cal T} = \frac{\nu}{(2\pi)^3}\int{\rm d}^3k \, k^2 \, n({\vec k},T)  \, ,
\end{align}
have been introduced.
The integral of $n({\vec k},T)$  over the available phase space at finite temperature $T$ gives the 
total density, $\rho$:
      \begin{equation}
      \rho=\frac{\nu}{(2\pi)^3}\int{\rm d}^3 k\,n(\vec k,T)\,,
      \label{eq:dens}
      \end{equation}
where $\nu$ is the degeneracy ($\nu = 4$ for spin and isospin saturated 
nuclear matter). This condition determines the chemical potential at a fixed external density.
The calculation of the single-particle energy and the density normalization condition defines 
a self-consistent process. 
Note, however, that due to the simple structure of the Skyrme interaction, 
self-consistency at the HF level is already achieved at the first iteration. 

Having fixed the chemical potential, the momentum distribution and the kinetic energy density are fixed. 
From these quantities, one can immediately compute the total HF energy per particle:
      \begin{align}
	e(\rho,T) &= \frac{\hbar^2}{2m^*} \frac{\cal T}{\rho} +
      \frac{3}{8}\rho\left[ t_0+\frac{1}{6}t_3\rho^\alpha \right]\,,
      \label{eq:eeta}
      \end{align}
and the entropy per particle:
\begin{equation}
s(\rho, T) =\frac {\nu}{(2 \pi)^3 \rho} \int d^3 k \left [ n(\vec k,T) \ln n(\vec k,T) + (1 -n(\vec k,T)) \ln (1-n(\vec k,T)) \right ] \, .
\end{equation}
In turn, these give access to the free energy, $f= e -Ts $, from which all the 
remaining thermodynamical properties can be computed. In addition to the chemical 
potential, $\mu$, the pressure, $P$, is also necessary to study the liquid-gas phase 
transition. One should take into account that this procedure is thermodynamically consistent,
in the sense that the chemical  potential extracted from the normalization of the density,
Eq.~(\ref{eq:dens}), coincides with the chemical potential derived from the free energy, 
$\mu = f+\rho\, (\partial f/\partial \rho)$, provided that the rearrangement term is properly included.

The Clausius-Clapeyron formula,
      \begin{equation}
      l=T\left(\frac{1}{\rho_g}-\frac{1}{\rho_l}\right) \left(\frac{{\rm d}P}{{\rm d}T}\right)_{coex}\,,
      \label{eq:clapeyron}
      \end{equation}
gives access to the latent heat per particle \cite{landau,huang}. This is expressed
in terms of the product of the temperature, the difference between volume
per particle of the two phases and the derivative of the pressure with respect to
the temperature along the coexistence curve. 
Alternatively, the latent heat can also be computed as the amount of heat that is 
needed, at a fixed temperature, to transfer one nucleon 
from the ordered (liquid) to the disordered (gas) 
phase. Since this process happens at constant chemical potential and pressure,
the heat change only involves the difference in entropy per particle of the two phases:
      \begin{equation}
      l=T(s_g-s_l)\,.
      \label{eq:latent}
      \end{equation}
For nuclear matter calculations, this formula is numerically more stable than the 
Clausius-Clapeyron one.  
Values of the vaporization specific latent heat for common 
liquids and gases are in the region of $10-5000$ kJ/kg when measured at the normal
boiling point. In contrast, as we shall see in the following, nuclear matter has a maximum 
specific latent heat of the order of $\sim 30$ MeV,
\emph{i.e.} $\sim 3\times10^{12}$ kJ/kg, which is orders of magnitude higher and among
the highest in nature (an exception being the quark-gluon plasma). The origin of this 
extremely large value can be traced back to the strong force, which binds the nucleons
tightly. 

A qualitative model that justifies these relatively high values and provides insight on the temperature dependence of the latent heat 
can be obtained using basic ingredients. From Eq.~(\ref{eq:latent}), we see that the latent heat can be computed from the difference of 
entropies times the temperature. 
On the one hand, let us take a classical approximation for the entropy of the gas,
\begin{align}
s_g = \frac{5}{2} - \frac{\mu_g}{T} \, .
\label{eq:s_gas_class}
\end{align}
This approximation should be valid because the gas coexistence density is, in general, quite low, 
which is also why interaction effects are neglected. On the other hand, the liquid phase is closer to 
saturation density and one could use the Sommerfeld expansion to compute its thermodynamical 
properties. The entropy in the liquid phase would then yield:
\begin{align}
s_l = \frac{\pi^2}{2} \frac{T}{\varepsilon_0} \, ,
\label{eq:s_Sommerfeld}
\end{align}
where $\varepsilon_0=\frac{\hbar^2 k_F^2}{2 m}$. We have ignored interaction effects (which would appear 
in the form of an effective mass) and we take $k_F$ at a reference density equal to 
saturation density. Within this approximation, the latent heat becomes:
\begin{align}
l = -\mu(T) + \frac{5}{2}T -  \frac{\pi^2}{2\varepsilon_0} T^2 \, .
\label{eq:l_approx1}
\end{align}
The chemical potential is the same for the liquid and the gas phases, due to the coexistence conditions (see Eq.~(\ref{eq:coex}) below). 
One can therefore use the Sommerfeld expansion in the liquid branch to find the corresponding chemical potential:
\begin{align}
\mu_l = \varepsilon_0 \left[ 1 - \frac{\pi^2}{12} \left( \frac{T}{\varepsilon_0} \right)^2 \right] + U_0\, .
\label{eq:mu_Sommerfeld}
\end{align}
Note that, again, we fix all densities to be at saturation and that $U_0$ is the attractive contribution needed 
so that $\mu_l(T=0)=e(\rho_0,T=0) \equiv -e_0$. Using Eq.~(\ref{eq:mu_Sommerfeld}) into Eq.~(\ref{eq:l_approx1}), one finds 
the following expression for the latent heat:
\begin{align}
l = e_0 + \frac{5}{2}T -  \frac{5 \pi^2}{12\varepsilon_0} T^2 \, .
\label{eq:l_approx}
\end{align}
Within this very crude model, the latent heat is a quadratic function of the temperature. Loosely speaking, 
classical effects (\emph{i.e.} those due to Eq.~(\ref{eq:s_gas_class})) tend to increase 
the latent heat linearly in temperature. Thermal effects arising from the degenerate expansion compensate 
this term and lead to a maximum, occurring at a temperature,
\begin{align}
T_{\textrm{max}} = \frac{3}{\pi^2} \varepsilon_0 \sim 11 \textrm{ MeV} \, .
\label{eq:t_max}
\end{align} 
The maximum value of the latent heat is then given by the sum of two terms:
\begin{align}
l_{\textrm{max}} = e_0 + \frac{15}{4 \pi^2} \varepsilon_0 \sim 30 \textrm{ MeV} \, .
\label{eq:l_max}
\end{align} 
In addition to the term due to the saturation energy, an equally important term (in sign and size) appears as a result of 
the competition between classical and degeneracy effects. This increases the value of the maximum 
latent heat substantially, up to around $30$ MeV. In the following, we will see that the predictions of this simple model are well fulfilled by Skyrme mean-field calculations of the liquid-gas phase transition.

\section{Results}

\begin{figure}
\begin{center}
      \includegraphics[scale=0.5,angle=-90]{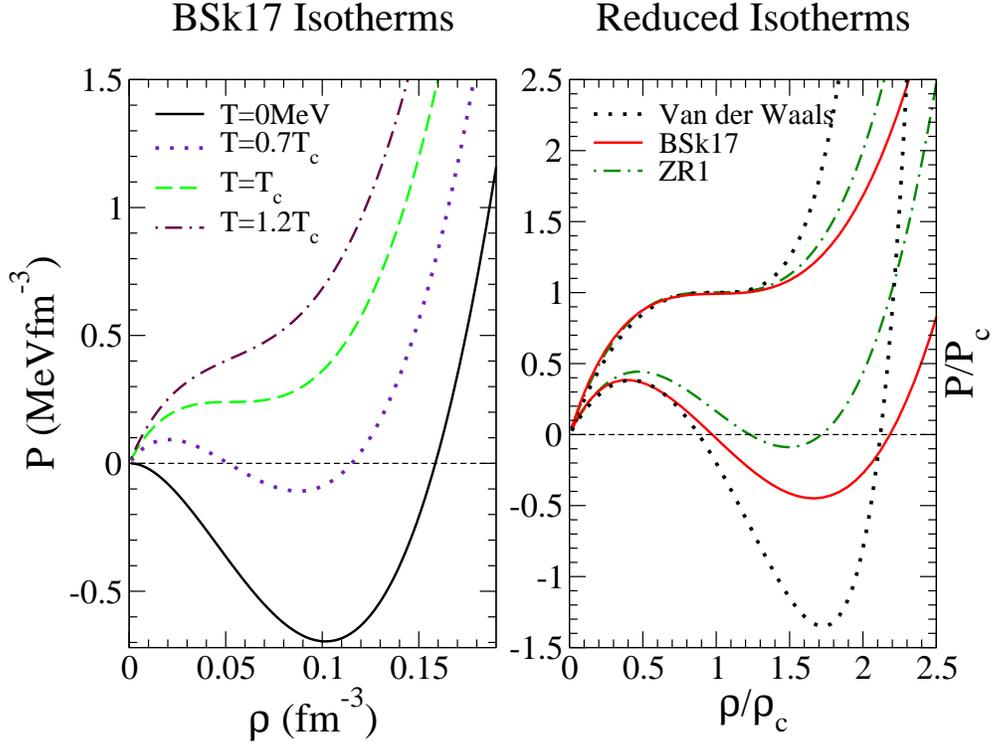}
      \caption{(Color online) Left panel: four different isotherms for the BSk17 Skyrme force, $T_c=15.6$MeV.
      Right panel: reduced isotherms at $T=T_c$ and $T=0.7 \, T_c$ for the BSk17 (solid lines) and ZR1 
      (dash-dotted lines) forces.  Isotherms of the Van der Waals model (dotted lines)
      are also shown for comparison.}
      \label{fig:bsk17}
\end{center}
\end{figure}

The liquid-gas phase coexistence and the critical properties of nuclear matter 
studied in a mean-field approximation with Skyrme forces are well-known \cite{rios10}.
Since the emphasis here is on the latent heat itself, we will only consider 
a few representative effective forces. We shall also briefly comment on 
some well known results on the phase coexistence before considering the latent heat. 
These will be helpful for our analysis. 
As a first example, we choose  the Skyrme force BSk17 \cite{goriely09}, which 
gives a very accurate description of the masses of nuclei all across the 
mass table.
Four pressure isotherms calculated with BSk17 are reported in the left panel of 
Fig.\ref{fig:bsk17}. These include the representative $T=0$ (solid line) 
and the critical $T=T_c$ (dashed line) isotherms. 
Two more isotherms, one below (dotted line) and one
above (dash-dotted line) the critical one, are also displayed. 
The isotherms show the well known shape associated to a liquid-gas phase transition. 
At $T=0$, the liquid at saturation density has zero pressure and, therfore, it is in equilibrium
with a zero density gas. As the temperature rises, the gas coexistence density
shifts to finite values and the coexistence region shrinks until the critical temperature is 
reached. Below $T_c$, all isotherms present a mechanically unstable region, where the
pressure decreases with density.
At $T_c$, phase coexistence is not possible anymore and the system vaporizes 
completely. In terms of isotherms, one finds that, for $T>T_c$, the pressure becomes a 
monotonically increasing function of density. 
For BSk17, the critical temperature turns out to be 15.6 MeV, similar to the 
$T_c$ of a wide range of modern Skyrme forces \cite{rios10}.

The behaviour of these isotherms is reminiscent of the Van der Waals equation of state
(EoS) for real gases which, unlike the ideal gas EoS, takes into account
the non-zero size of the molecules of the gas \cite{huang}. The excluded volume
has a repulsive effect, in contrast to the attractive inter-molecular force.
This causes the formation of regions of instability which result in a 
liquid-gas phase transition. 
To appreciate the similitudes and differences between the Van der Waals and the
self-consistent mean-field EoS, 
we plot in the right panel of Fig.~\ref{fig:bsk17} the reduced
isotherms  ($P/P_c$ vs. $\rho/\rho_c$ ) at $T=0.7 \,T_c$ and at $T_c$. In addition to
the Van der Waals case, we consider the EoS obtained with BSk17 and with the ZR1 
\cite{jaqaman84} forces. The latter has the highest critical temperature
($T_c=22.98$ MeV) among the large set of effective forces analyzed in Ref.~\cite{rios10}.

By studying the pressure in reduced units ($P/P_c$, instead of $P$) we expect to 
highlight possible resemblances between several EoS. The Van der Waals model describes
the EoS of monatomic 
gases and liquids within a reasonable distance above and below their critical point
\cite{huang}.
Moreover, according to the principle 
of corresponding states, if we measure pressure, volume and temperature in units of $P_c,\rho_c$ and $T_c$, 
the EoS becomes universal, \emph{i.e.} it is the same for a wide range of substances \cite{huang}.
One might wonder whether there is something like a principle of corresponding states
for EoS derived from different mean-fields. The right panel of Fig.~\ref{fig:bsk17}
provides an insight into this matter. For $T=0.7 \,T_c$, beyond the gas phase 
there is a relatively important disagreement between all the reduced EoS. It is only close to the 
critical point that the EoS coincide. 
This is a consequence, as we will see later,  of the fact that the Van der Waals and the self-consistent mean-field 
models have the same critical exponents \cite{rios10}.

\begin{figure}
      \begin{center}
      \includegraphics[scale=0.4,angle=-90]{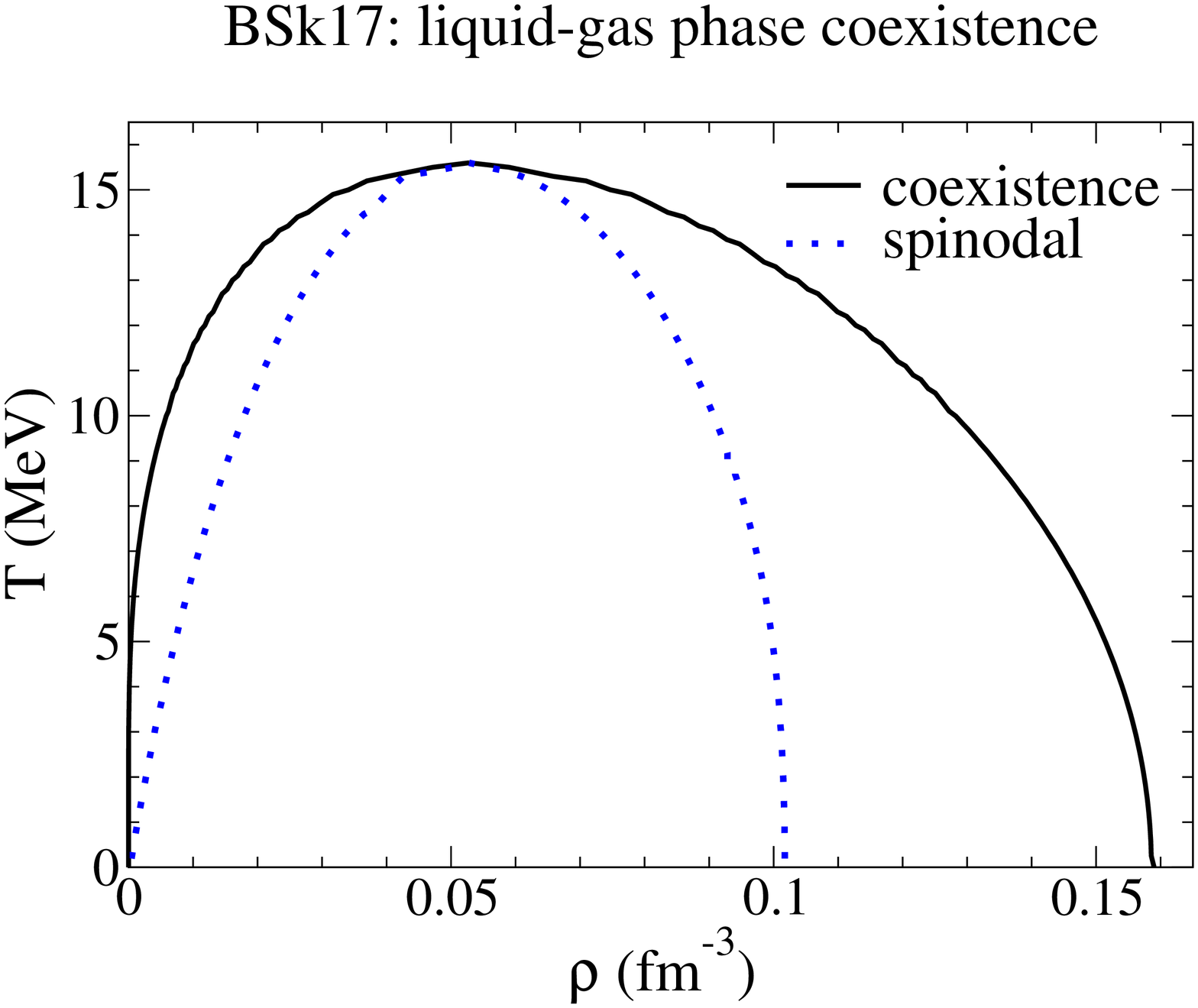}
      \caption{(Color online) Coexistence line (solid) and spinodal line (dotted) obtained 
      from the BSk17 force.}
      \label{fig:liquidgas.bsk17}
      \end{center}
\end{figure}

When nuclear matter is heated up, a phase coexistence develops between a relatively
high-density liquid phase and a low-density gas. In order for such equilibrium to
exist, the chemical potential and the pressure of the two phases should be equal:
     \begin{equation}
      \mu(\rho_g,T)=\mu(\rho_l,T)\,,\quad P(\rho_g,T)=P(\rho_l,T)\,.
      \label{eq:coex}
      \end{equation}
At a given temperature, $T$, the solution of this set of equations provides 
a density couple, $(\rho_g,\rho_l)$, which defines the coexistence densities of the two
phases. In Fig.~\ref{fig:liquidgas.bsk17},
we plot the coexistence phase diagram (solid line) in the $(\rho,T)$ plane for the BSk17 force. 
As it has already been mentioned, 
at $T=0$ the liquid at saturation density is in coexistence with a zero-density gas. 
The gas coexistence density grows and the liquid coexistence density decreases 
as temperature rises. The densities of the gas and liquid phases
join at the critical point, above which the liquid-gas phase coexistence disappears.
Together with the coexistence line, we also plot the spinodal line (dotted line), 
which marks the boundary between thermodynamically 
stable and unstable states of matter. 

\begin{figure}
      \begin{center}
      \includegraphics[scale=0.5,angle=-90]{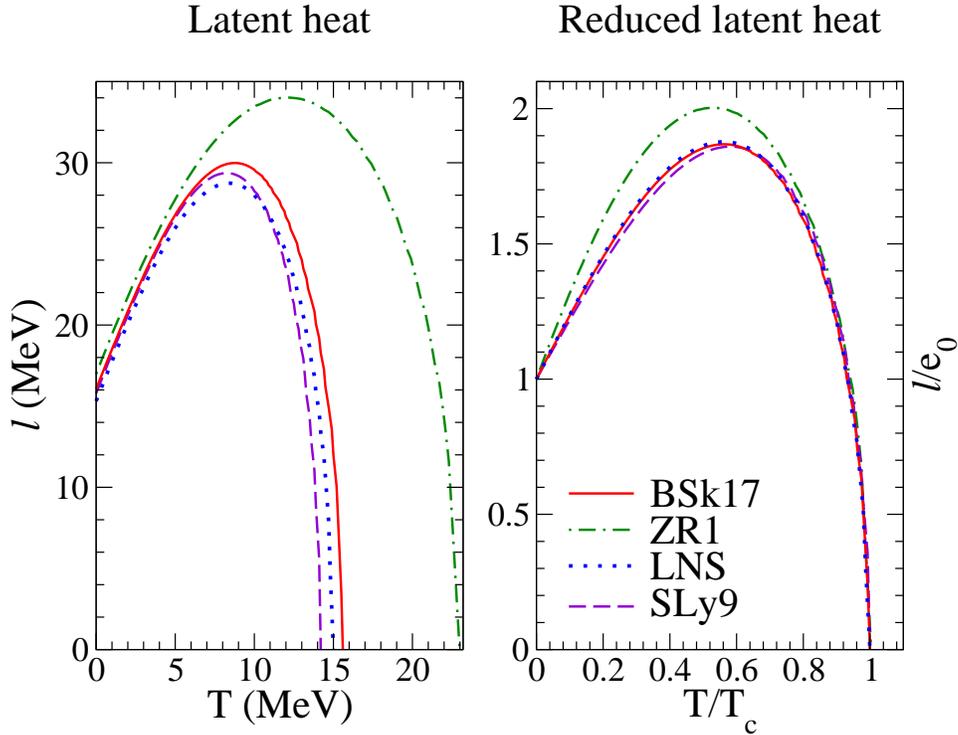}
      \caption{(Color online) Left panel: latent heat using four different Skyrme force parametrizations. Right panel:         
      latent heat in reduced dimensionless units using the same forces. }
      \label{fig:lh}
      \end{center}
\end{figure}

The latent heat corresponding to the liquid-gas phase transition obtained with BSk17 using
Eq.~(\ref{eq:latent}) is reported in the left panel of Fig.~\ref{fig:lh} (solid line). 
One can observe that the latent heat has a characteristic bump shape as a function of temperature. In the $T=0$ limit, 
$l$ has a finite value. For BSk17 (solid line), it grows with temperature up to 
a well defined maximum at $T=8.7$ MeV (with $l_{\textrm{max}}= 29.9$ MeV) 
and then sharply goes to zero at $T=T_c$.
To appreciate better the dependence of the latent heat of nuclear matter on the
different Skyrme parameterizations, we have also performed calculations of the
coexistence line and latent heat for various other forces. 
In addition to the already mentioned BSk17 and ZR1, 
we have chosen two more interactions: SLy9 \cite{chabanat,chabanat97}, which
incorporates, by construction, the behavior of a microscopically derived EoS of neutron matter \cite{wiringa88} 
and has a very low critical temperature \cite{rios10}; 
and LNS, that reproduces
the properties of BHF calculations of nuclear matter \cite{lombardo06}.
The results obtained for their latent heats are also reported in the left panel of Fig.~\ref{fig:lh}.
Note that all forces produce a latent heat with a similar qualitative behavior that, as we shall 
discuss below, can be understood from basic principles.

Let us start the discussion with the zero temperature limit of the latent heat. As mentioned 
previously, $l$ measures the heat that needs to be
provided to the fluid to transfer a nucleon from the liquid to the gas phase.
Since the equilibrium coexistence density of the gas  asymptotically goes to 
zero when taking the $T \to 0$ limit, $l$ becomes
the amount of heat needed to \emph{extract} a particle from the system in that limit. 
In other words, for $T \to 0$, $l \to -\mu$. 
At the saturation density of nuclear matter, $\rho_0$, one actually finds 
$\mu=-e_0$ and, consequently, the latent heat is equal in absolute
value to the saturation energy, $l=e_0$, when getting close to $T=0$. 
This result is independent of the interaction or the many-body
approximation used to describe nuclear matter, as long as thermodynamical
consistency is fulfilled. 
This suggests that, if one wants to obtain interaction-independent results
for the latent heat, $l$ might be normalized to $e_0$ (see right panel of Fig.~\ref{fig:lh}).

We observe that for all Skyrme parametrizations the qualitative behaviour of the latent heat
is very similar (left panel of Fig.~\ref{fig:lh}).
The latent heat matches
the value of the binding energy at $T=0$, then rises for small
temperatures. The initial rise is linear and the slope seems to 
be independent of the Skyrme parametrization. As we shall see below, the slope
of $l$ close to $T=0$ is a model-independent feature that can be understood from
fundamental arguments. Further up in temperature, $l$ 
reaches a maximum and then drops to zero at the critical point,
where the difference between the liquid and gas phases disappears.
The results with this limited set of forces seem to indicate that
the position and magnitude of the maximum in $l$ 
depend on the specific value of the critical temperature. Broadly speaking, 
higher values of $T_c$ shift the position and height of 
the maximum to larger values. This is particularly clear for the case of the 
ZR1 mean-field, which has the largest critical temperature. A calculation with a wider
set of mean-fields, not shown here for simplicity, confirms this tendency. 

A plot of the latent heat in reduced units, $l/e_0$, as a function of the reduced
temperature, $T/T_c$, is presented in the right panel of Fig.~\ref{fig:lh}. Similar to the right
panel of Fig.~\ref{fig:bsk17}, the reduced plot is helpful in highlighting the dependence of the 
latent heat on the different EoS. In general terms, we observe that the large dependence
on the mean-field is eliminated to a large extent in the dimensionless plot. 
Close to $T=0$, the 
linear slope of $l$ is changed due to the fact that different mean-fields have different 
saturation energies, $e_0$. Yet, near the
maximum, the reduced latent heats show a much smaller deviation compared 
to the absolute ones.
For all Skyrme forces, the latent heat tends to peak within a limited region of 
temperatures,
$T/T_c \sim 0.5-0.6$. Moreover, the peak is also quite narrowly distributed
around the value $l_{max}/e_0 \sim 1.7-2$, which suggests that the latent heat is 
more determined by
thermal correlations than by effective forces. Finally, as the temperature reaches the 
critical value, the latent heat falls to zero with a very similar temperature dependence for
all forces. This identical
behavior can be explained in terms of critical exponents  [see Eq.~(\ref{eq:critexp})]. 
As it will be shown later, within the mean-field approximation, the critical exponents of all 
latent heats are the same close to the critical point.

\begin{figure}
\begin{center}
      \includegraphics[scale=0.4,angle=-90]{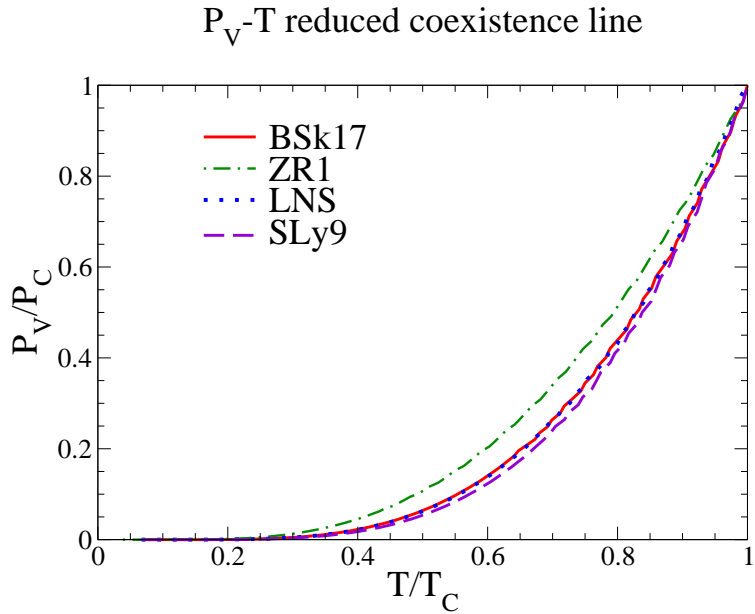}
      \caption{(Color online) Vapor pressure versus temperature in reduced dimensionless units
       obtained from the equilibrium conditions Eq.~(\ref{eq:coex}).}
      \label{fig:p_t_red}
\end{center}
\end{figure}

As mentioned in the previous section, the Clausius-Clapeyron formula requires the evaluation of the
derivative of the pressure with respect to the temperature along the coexistence line. 
The pressure along coexistence is commonly referred to as the vapor pressure. 
Its behavior in reduced dimensionless units, 
$P_v/P_c$, along the phase transition is shown in 
Fig.~\ref{fig:p_t_red} for the four mentioned Skyrme
parametrizations. The vapor pressure is a well behaved function of the temperature that grows from
zero to $P_c$ as temperature increases from zero to $T_c$. For low temperatures, $P_v$ rises very 
slowly and at $T=0.5 \, T_c$ it is only $10$ \% of $P_c$. 
Above this temperature, a steady increase brings the vapor pressure very rapidly up to 
$P_c$. Note that the last portion of this increase is basically linear, in accordance to 
Eq.~(\ref{eq:picoex}) below.
We have checked the numerical and thermodynamical consistency between the values of the latent heat
obtained with the Clausius-Clapeyron equation (Eq.~(\ref{eq:clapeyron})) and those given by the difference 
of entropies of the gas and the liquid phase (Eq.~(\ref{eq:latent})).

\begin{figure}
\begin{center}
      \includegraphics[scale=0.4,angle=-90]{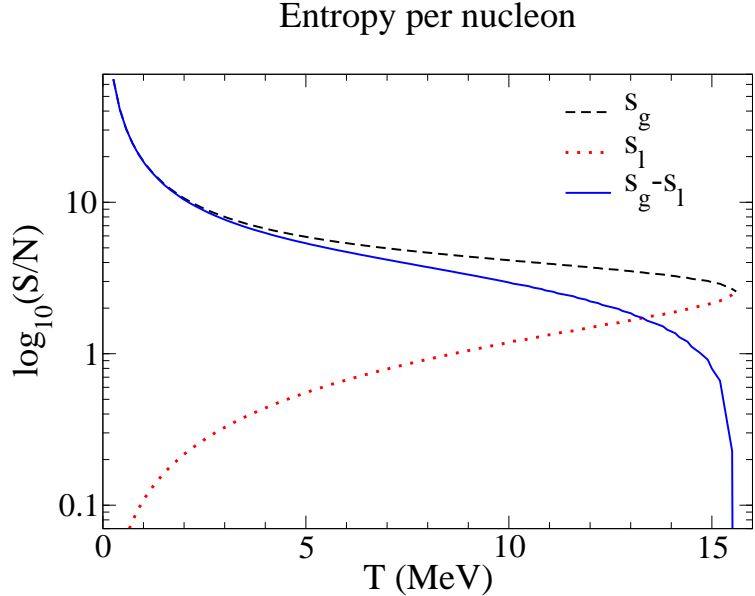}
      \caption{(Color online) Logarithmic plot of the difference of gas and liquid entropies (solid line), 
      entropy of the liquid (dashed line) and 
      entropy of the gas (dotted line) along the coexistence line using the BSk17 Skyrme force.}
      \label{fig:entrogl}
\end{center}
\end{figure}

A basic ingredient in the evaluation of the latent heat is the difference between the 
entropy of the gas and that of the liquid phase [see Eq.~(\ref{eq:latent})]. 
This difference, which has to be evaluated along the coexistence line, is
shown in Fig.~\ref{fig:entrogl} for the BSk17 interaction. Entropies are plotted in a logarithmic 
scale.  
The difference between the entropy of the gas and that of the liquid is always a positive 
quantity that decreases with temperature and goes to  
zero as the system reaches the critical temperature. 
The low-temperature limit of this difference can be studied analytically and
will be discussed in the next section. Note that, within this limit,
the difference between the gas and liquid
entropy is largely dominated by the entropy of the gas. 
Also notice that the variation of the liquid entropy, which should be zero at $T=0$, 
is much smaller than the variation of the entropy associated to the gas phase (note the 
logarithmic scale). 
While the liquid entropy along the coexistence line is an increasing function of temperature,
the gas entropy is a decreasing one. Their difference, however, is dominated by the
gas entropy and becomes a decreasing function of temperature.
As a matter of fact, in the context of classical gases, it is customary to neglect
the liquid contribution, since its entropy (or, in terms of Eq.~(\ref{eq:clapeyron}), its inverse 
volume) is negligible with respect to the gas one \cite{landau}. In nuclear matter, this 
approximation would only be valid up to 
$T/T_c \sim 1/3$, which would lead to a maximum error in the calculation of the latent heat of about 10\%. 
For higher temperatures the error induced by this approximation would already reach the 30\%
for the maximum of the latent heat.
Finally, let us stress that the maximum that appears in the latent heat at intermediate 
temperatures is 
a subtle result, arising from the product of the (linearly increasing) temperature times the 
(decreasing function of $T$) difference of entropies. 

\section{Analytical limits of the latent heat}
\subsection{Low temperature behaviour}

As we have already discussed, all Skyrme forces produce a qualitative similar behaviour of 
the latent heat. 
The value in the zero temperature limit ($l=e_0$) is well understood from basic arguments. 
In this limit, the liquid phase is at the saturation density and therefore
has zero pressure. The heat needed to transfer a nucleon from the homogeneous liquid 
nuclear matter phase to the zero density gas (vacuum) 
is just the chemical potential of the liquid which, at saturation, coincides with the binding energy per particle. 
In this subsection we will show analytically that 
this intuitive value is the result of a delicate balance. 

In our derivation, we will not only evaluate the value of $l$ at $T=0$, but also its 
derivative as a function of temperature at this point. This derivative is 
basic in understanding the 
existence of a maximum in $l$. The argument goes as follows: at $T=0$, the latent heat
is finite and positive. As $T \to T_c$, 
however, the latent heat must go to zero, 
since the liquid-gas phase transition disappears. 
The existence of the maximum is therefore necessarily related 
to the way $l$ departs from its $T=0$ value. If the slope at $T=0$ is positive, the function 
will first increase with $T$ and, in order to reach the zero value at $T_c$, at least one 
maximum will have to develop at some intermediate 
temperature. We shall also show that the slope of $l$ at $T=0$ is not only 
positive, but model-independent.

Let us consider the limit of low temperatures of the latent heat. 
As shown in Fig.~\ref{fig:liquidgas.bsk17}, as $T$ approaches to zero, the dense liquid 
phase is in equilibrium with a very low-density gas. 
Due to its diluteness, we shall assume that the gas is
in the classical regime and interactions are not relevant anymore. The thermodynamical properties of 
the gas are then given by ideal gas expressions \cite{huang} and they can
be computed analytically for a given gas density, $\rho_g$, and temperature, $T$. The 
pressure is simply given by
\begin{equation}
P = \rho_g T \, ,
\end{equation}
while the gas density can be obtained in terms of its chemical potential by using the 
following expression:
\begin{equation}
\rho_g = \nu \left ( \frac {m}{2 \pi \hbar^2 }\right )^{3/2}  T^{3/2} e^{\mu_g/T} \, .
\label{eq:rhogas}
\end{equation}
If we substitute the previous expression in the pressure, we find the vapor pressure of
nuclear matter:
\begin{equation}
P_v (T) = \nu \left ( \frac {m}{2 \pi \hbar^2}\right )^{3/2} T^{5/2} e^{\mu_g/T} \, .
\label{eq:gaspres}
\end{equation}
Notice that the information on the interactions in nuclear matter is contained in the chemical 
potential, $\mu_g$, which is the same for the liquid and the gas according to the 
equilibrium relation,
Eq.~(\ref{eq:coex}). In the liquid branch, as $T$ approaches zero, the chemical potential
tends to the energy per particle, and one has $\mu_g = \mu_l \to -e_0$.

According to the Clausius-Clapeyron formula, Eq.~(\ref{eq:clapeyron}), to calculate $l$ we 
need the temperature derivative of the vapor pressure along the coexistence curve. Using 
Eq.~(\ref{eq:gaspres}), this derivative becomes:
\begin{equation}
\frac {dP_v}{dT} = \nu \left ( \frac {m}{2 \pi \hbar^2 }\right )^{3/2} \sqrt{T}
\left [ -\mu_g + T \left( \frac {5}{2}  - 
\frac{{\rm d} \mu_g}{{\rm d} T} \right) \right ] e^{\mu_g/T} =
\frac{\rho_g}{T}
\left [ -\mu_g + T \left( \frac {5}{2}  - 
\frac{{\rm d} \mu_g}{{\rm d} T} \right) \right ] 
 \, .
 \label{eq:dervapres}
\end{equation}
All temperature derivatives are to be taken as derivatives along the 
coexistence line. Because of
the equilibrium condition in Eq.~(\ref{eq:coex}), the chemical potential can be computed
from the
liquid one which, close to saturation, should be that of a degenerate Fermi gas. The
Sommerfeld expansion then guarantees that the temperature dependence of $\mu_l$ is
quadratic in $T$ \cite{ashcroft}. Consequently, the temperature derivative in the last term
$\frac{{\rm d} \mu_g}{{\rm d} T} \sim \mathcal{O}(T)$ and thus it can be 
neglected in the following considerations. 
Note that the low temperature limit of the previous expression, Eq.~(\ref{eq:dervapres}), 
is zero. However, when
the latent heat is considered in the zero temperature limit, we need to explicitly take into account
the prefactors, whose cancelation leads to:
\begin{eqnarray}
\lim_{T \rightarrow 0} l(T) = 
\lim_{T \rightarrow 0} T \left ( \frac {1}{\rho_g}- \frac {1}{\rho_l} \right ) \frac {d P_v}{d T} 
&=& \lim_{T \rightarrow 0} T \frac {1}{\rho_g} \frac {d P_v}{d T}
=   - \mu_g = e_0 \, .
\label{eq:latent_T0}
\end{eqnarray}
When taking the limit, we have considered that the term containing the liquid density goes to zero 
because the gas density present in the derivative of the vapor pressure 
(see Eq.~(\ref{eq:dervapres})) goes to zero in this limit, and that the linear terms in temperature within 
the brackets in Eq.~(\ref{eq:dervapres}) are subleading.

Let us now compute the derivative of $l$ close to zero temperature:
\begin{eqnarray}
\frac {dl}{dT}= \left ( \frac {1}{\rho_g} - \frac {1}{\rho_l} \right ) \frac {dP_v}{dT} 
	+ T \left ( \frac {1}{\rho_g} - \frac {1}{\rho_l} \right ) \frac {d^2 P_v}{dT^2}
	+ T \frac {dP_v}{dT} \frac {d}{dT} \left ( \frac {1}{\rho_g} - \frac {1}{\rho_l} \right )\,.
\end{eqnarray}
The $T \to 0$ limit of the first term is obtained from Eq.~(\ref{eq:dervapres}). The second
term involves the second derivative of the vapor pressure, which is also easily computed
from Eq.~(\ref{eq:dervapres}) and yields:
\begin{equation}
T \left ( \frac {1}{\rho_g} - \frac {1}{\rho_l} \right ) \frac {d^2 P_v}{dT^2} \sim
-\frac{1}{T^2} \left( \mu^2 - 4 \mu T + \frac{15}{4} T^2 \right) \, .
\end{equation}
The third term involves derivatives of the gas density, Eq.~(\ref{eq:rhogas}), while the liquid
density is neglected:
\begin{equation}
T \frac {dP_v}{dT} \frac {d}{dT} \left ( \frac {1}{\rho_g} - \frac {1}{\rho_l} \right )\sim
\frac{1}{T^2} \left( \mu^2 - 3 \mu T + \frac{15}{4}  T^2  \right) \, .
\end{equation}
Collecting the different contributions, one gets :
\begin{eqnarray}
\lim_{T\rightarrow 0} \frac {dl}{dT} = \left [\frac {5}{2} - \frac {\mu}{T} \right ] 
+ \left [ \frac {\mu^2}{T^2} - \frac {3\mu}{T} + \frac {15}{4} \right] 
- \left [ \frac {\mu^2}{T^2} - \frac {4\mu}{T} + \frac {15}{4} \right] = \frac {5}{2}
\label{eq:derlatent}
\end{eqnarray}
The derivative of the latent heat with respect to the temperature
in the limit $T\rightarrow 0$
is therefore independent of the interaction. Moreover, it is positive and equal to $5/2$.

\begin{figure}
\begin{center}
      \includegraphics[scale=0.4,angle=-90]{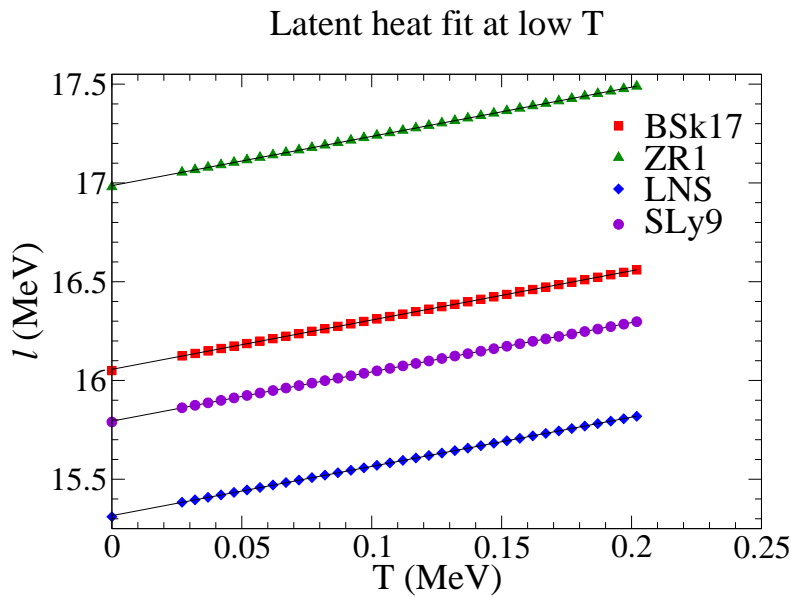}
      \caption{(Color online) Latent heat versus temperature for very low $T$. Symbols represent 
      self-consistent mean-field results and lines their respective linear regression fits.}
      \label{fig:latent0}
\end{center}
\end{figure}

In Fig.~\ref{fig:latent0} we show a numerical proof of Eq.~(\ref{eq:derlatent}). For the
different Skyrme force parametrizations considered in Fig.~\ref{fig:lh}, we focus on the low-temperature 
behaviour of the latent heat (symbols). 
The slopes for the linear regression fits (lines) of these numerical data 
are in very good agreement with the value $5/2$ up to two significant digits. 

This result is not only valid regardless of the effective interaction, 
but it is also valid no matter which many-body approximation is considered.
The only assumption that has been made is that the
gas equilibrium density enters the classical regime as the temperature decreases. 
Consequently, 
one should get the same result in approaches that go beyond the HF
approximation. It is also important to note that this result is independent of the
system under study and therefore should be generically valid for the liquid-gas phase 
transition of any extended normal fermionic system. As discussed previously,
the positiveness of this derivative necessarily implies that a maximum in the 
latent heat must develop. As a result, we expect a maximum in the latent heat of 
any normal fermionic system that presents a liquid-gas phase transition.

\subsection{Critical behaviour}

\begin{figure}
\begin{center}
      \includegraphics[width=0.5\linewidth]{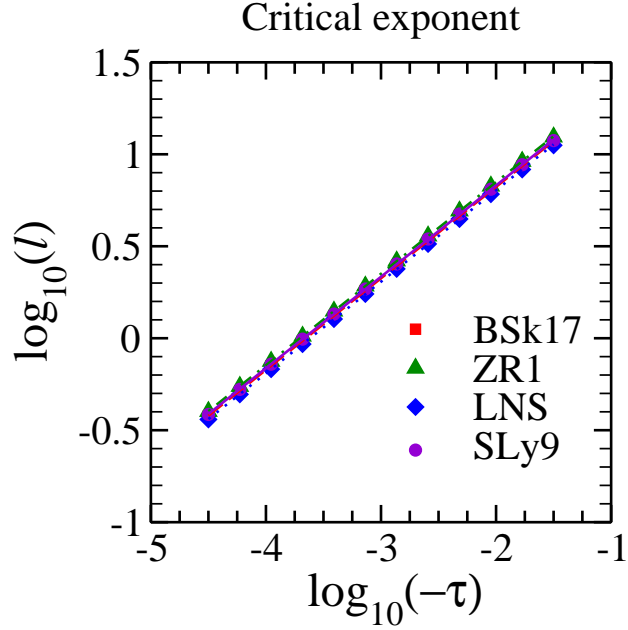}
      \caption{(Color online) Logarithmic plot of the latent heat versus the reduced temperature. 
      The slope of the different curves gives the critical exponent. Symbols represent self-consistent mean-field 
      results and lines their respective linear regression fits.}
      \label{fig:c_exponent}
\end{center}
\end{figure}

Critical exponents characterize the properties of phase transitions
\cite{huang,landau,kadanoff67}. 
Close to the critical point, all the thermodynamical properties can
be described in terms of a handful of exponents. Their knowledge facilitates the 
understanding of the properties of systems close to criticality. The latent heat is not an 
exception
and it can also be described, close to the critical point, in terms of a critical exponent. 
As shown in Ref.~\cite{landau}, the critical exponent for $l$ is the same as that of the 
order parameter.

This result can be understood in terms of the mean-field theory of fluctuations 
as follows. Consider the reduced pressure, 
$\pi = (P-P_c)/P_c$, close to the critical point. Expanding to first order in the
reduced temperature, $\tau=(T-T_c)/T_c$, and to third order in the reduced density,
$\eta=(\rho-\rho_c)/\rho_c$, one finds:
\begin{align}
	\pi = a \tau + b \tau \eta + c \eta^3 + c' \tau \eta^2 \,. 
	\label{eq:pi}
\end{align}
Note that the conditions to find the critical point,
\begin{align}
	\frac{\partial \pi}{\partial \eta} =  	\frac{\partial^2 \pi}{\partial \eta^2} =0 \, ,
\end{align}
have already been used to eliminate possible explicit linear and quadratic terms in $\eta$. 
To find the pressure at phase coexistence, one can use Eq.~(\ref{eq:coex}) in 
Eq.~(\ref{eq:pi}) to find the symmetric gas and liquid coexistence points. One can then
show that the coexistence pressure is given by:
\begin{align}
	\pi |_{coex} = a \tau \, .
	\label{eq:picoex}
\end{align}
Using the Clausius-Clapeyron relation, the latent heat close to the critical 
point becomes:
\begin{align}
	l = \frac{a P_c}{2 \rho_c} \left( \rho_l-\rho_g \right) (1+\tau)  \sim ( -\tau )^\beta\, .
	\label{eq:critexp}
\end{align}
In this expression, we have considered that the difference in densities
is the order parameter of the phase transition and that the latter is governed
by the $\beta$ critical exponent, $ \rho_l-\rho_g \sim (- \tau)^\beta$. As any
mean-field theory \cite{rios10,huang}, self-consistent
Hartree-Fock calculations of the nuclear matter liquid-gas phase transition yield a
critical exponent $\beta = 1/2$.

We have checked numerically that the critical exponent of $l$ is indeed $\beta = 1/2$. 
Fig.~\ref{fig:c_exponent} shows a logarithmic plot of $l$ versus the reduced critical 
temperature (symbols). Linear regression fits have been performed and are shown
with lines. The correlation coefficients are close to $1$ to within at least four digits in 
all cases. Such a linear behavior of the data confirms the scaling of $l$ 
with $\tau$. 
The slopes of these lines have also been extracted and agreement with the $\beta$
derived from the coexistence line is good up to the third digit. 
To our knowledge, this is the first time that the critical exponent of the
latent heat for nuclear matter is computed and that its equivalence to $\beta$ is 
confirmed numerically in the framework of a Hartree-Fock mean-field theory. 
However, one should keep in mind the limitations of the mean-field theory for the calculation of the critical
exponents. 

\section{Summary and conclusions}

In this work, we have have analyzed in detail the latent heat of the liquid-gas phase transition 
of symmetric nuclear matter. 
The latent heat describes the amount of heat needed to transfer one nucleon from the 
liquid to the gas phase, and it can be used as a further characterization of the phase 
transition. We have been motivated by experimental results which suggest that, in 
the phase transition occurring in nuclear multifragmentation collisions, a latent
heat of around $l \sim 2-8$ MeV can be observed \cite{pochodzalla97,dagostino00,bonnet09}. 

We have used self-consistent Hartree-Fock calculations at finite temperature to obtain
numerical results. Four different characteristic mean-fields have been used to determine
the inherent mean-field dependence of the results. 
The qualitative behavior of the latent heat as a function of temperature is very similar for 
all effective interactions. Within the Hartree-Fock approximation, the latent heats derived 
from different mean-field parametrizations fall within a narrow band when the temperature is 
scaled by $T_c$ and $l$ is scaled by $e_0$. 

In the $T \to 0$ limit, the latent heat coincides with the binding
energy per particle at saturation density. The latent heat can also be
computed from the difference in entropies between the gas and the liquid phases. We have
seen that the gas entropy dominates over the liquid one in a wide range of temperatures.
At finite but low temperatures,
the latent heat rises linearly with temperature. A careful analysis shows that the slope
of this linear trend is $5/2$, regardless of the mean-field parametrization or the many-body
approximation. We have confirmed this trend with numerical finite temperature 
HF calculations. This model-independent result is valid for all normal fermionic
systems. To our knowledge, it is the first time that this result is discussed. 

Knowing that 
a)  $l$ is positive and finite at $T=0$, b) it has a positive slope near $T=0$ and c) 
it goes to zero at $T=T_c$, necessarily implies the existence of a maximum in the latent
heat. 
Mean-field numerical calculations suggest that this maximum is located around 
$T \sim 0.6 T_c$ and that $l_{max} \sim 1.8 e_0$. While the exact position of the 
well-defined maximum can depend on the effective force or on the many-body approach 
used to describe nuclear matter, its existence is guaranteed independently of the mean-field
parametrization or the many-body approximation. Moreover, this result should be valid for 
any normal fermionic system that presents a liquid-gas phase transition.

It is not particularly easy to understand physically why the latent heat should present 
a maximum as a function of temperature. One might interpret the presence of this 
maximum as a manifestation of the underlying NN interaction. At low temperatures,
the latent heat rises because more work is needed to break the attractive part of
the strong interaction when transferring a nucleon from the more structured liquid to the 
gas phase. At higher temperatures, the thermal motion of the particles would provide
most of the work and thus less latent heat
is needed to break the bonding between particles in the liquid. Here, we have given an
analytical demonstration of the existence of the peak. 

Concerning the behavior of the latent heat near the critical point, we have shown numerically 
that it can be described in terms of  a single critical exponent. As predicted by the theory
of phase transitions, this critical exponent is the same as the one associated to the order
parameter. In the case of Hartree-Fock calculations in nuclear matter, $\beta=1/2$, and
there is a very good numerical agreement between both exponents. All in all, the
latent heat can be characterized in the low, intermediate and (close to) critical
temperature regimes from very basic principles. 
It is precisely this generic nature which might motivate the use of the latent heat as a
tool in analyzing the liquid-gas phase transition in other normal fermionic systems.

We are well aware that a connection between experimental observations of 
multifragmentation collisions
and theoretical results of homogenous nuclear matter is not at all transparent. 
Finite size effects play a capital role in determining the thermodynamical properties
of nuclei and the latent heat is not an exception \cite{lee97}. 
Nuclear matter values at its maximum
suggest that the latent heat is up to $10$ times higher than that extracted from
different experimental analysis \cite{pochodzalla95,dagostino00,bonnet09}.
Even at the theoretical level, it is
not clear how to define a liquid and a gas phase in a self-confined system 
\cite{bonche84}. 
Nevertheless, one might naively expect that the appearance of a maximum in the latent heat 
might have a substantial suppression effect on the yields of light particles. 
We hope that the present study, where we have highlighted model-independent
and basic arguments for the latent heat, will encourage further experimental 
analysis of the latent heat in multifragmentation collisions.

\section{Acknowledgements}

This work has been supported by Grants No. FIS2008-01661 (Spain), and No. 2009-SGR1289 
from Generalitat de Catalunya, 
a Marie Curie Intra European Fellowship within the 7$^{th}$ 
Framework programme and STFC grant ST/F012012, 
FCT (Portugal) under project CERN/FP/109316/2009, 
and by COMPSTAR, an ESF Research Networking Programme.
The authors are very grateful to professor A. Planes for useful and stimulating discussions.

\bibliographystyle{apsrev}
\bibliography{biblio}

\end{document}